\documentstyle[sprocl,12pt]{article}

\def\bea{\begin{eqnarray}}
\def\eea{\end{eqnarray}}

\begin{document}

\title{The harmonic gauge condition in the gravitomagnetic equations}
\author{J.-F. Pascual-S\'anchez}
\address{Dept. Matem\'atica Aplicada Fundamental,
Secci\'on Facultad de Ciencias,\\ Universidad de Valladolid,
47005, Valladolid,\\ Spain\\ E-mail: jfpascua@maf.uva.es}
\maketitle

\abstract{It has been asserted in the literature that the  analogy
between the linear and first order slow motion approximation of
Einstein equations of General Relativity (gravitomagnetic
equations) and
 the Maxwell-Lorentz equations of electrodynamics
breaks down if the gravitational potentials are time dependent.
 In this work,
 we show that this assertion is not correct and it has
 arisen from an incorrect limit of the usual
harmonic gauge condition, which drastically changes the physical
content of the gravitomagnetic equations. }


\setlength{\parskip}{4mm}

\section{Introduction}
It has been asserted in [1] and also quoted in [2], that the
analogy between the linear and first order slow motion
approximations of Einstein equations of general relativity (GR)
and Maxwell-Lorentz equations of electrodynamics, breaks down if
the gravitational potentials are time dependent. This assertion
comes, in our opinion, from the use of an incorrect limit of the
usual harmonic gauge condition, which implies the absence of the
gravitomagnetic induction term in the Faraday-like equation for
the gravitoelectric field ${\bf{g}}$. On the other hand, the
presence of the gravitomagnetic induction term is essential in
order to consider the possibility of existence of a spin dynamo.
This possibility, in spite of the results exposed in [3], merits
further study when the fluid equations are combined with the
gravitomagnetic equations.

 In this paper, we first review the linear approximation of GR in
 terms of the gravitational potentials and, in second place, in
 terms of a object in which only appear
   first derivatives of the gravitational potentials. In both cases,
  the harmonic gauge condition is imposed. After, we shall show
  that from the latter equations, when also the first order slow motion
  approximation is considered and even when the gravitational
  potentials are time dependent, one obtains gravitomagnetic
equations, which are completely analogous to the Maxwell-Lorentz
equations of electrodynamics. The only difference is that the sign
of the sources is reversed, to account for the attractive nature
of gravity between masses.

\section{Linear Einstein equations in terms of the gravitational potentials}
The procedure for linearizing the geometrical left hand side part
of Einstein's field equations for weak gravitational fields,
 is included in almost all
texts on GR, see for instance [2,4]. Starting with Einstein field
equations:
\begin{equation}\label{ein}
R_{a b}-\frac{1}{2}g_{a b}R= -\frac{8\pi G}{c^{4}}%
T_{a b}
\end{equation}
We retain the universal constants $G$ and $c$ throughout this
work, in order to know the different orders of the approximations.
If the gravitational field is weak, then the metric tensor can be
approximated by
\begin{equation}\label{1}
g_{a b}\simeq \eta _{a b}+h_{a b}
\end{equation}
where latin indexes $a, b = 0,1,2,3$ and $\eta _{a b }=\left(
+1,-1,-1,-1\right) $ is the flat Minkowski spacetime metric
tensor, and $h_{a b}$ is the gravitational perturbation to the
flat metric. When gravity is weak, the linear approximation to GR
should be valid. Thus, we now retain first order terms only, i.e.,
we neglect non-linear terms in $h_{a b}$ and its derivatives in
the Einstein equations, so $R_{a b}$ and $R$ are also correct to
first order. The Ricci tensor can be calculated from the tensorial
contraction of the Riemann tensor and the curvature scalar can be
calculated from the contraction of the Ricci tensor and read
\begin{equation} \label{Ricci}
R_{a b}\simeq -\frac{1}{2} \Box h_{a b}
\end{equation}
\begin{equation} \label{scal}
R\simeq \eta ^{a b}R_{a b}= -\frac{1}{2}\Box h ,
\end{equation}
where $\Box$ is the D'Alembertian operator and to obtain
(\ref{Ricci}) and (\ref{scal}) we have chosen our coordinate
system as harmonic, so that we have chosen the following harmonic
or de Donder gauge condition
 (four components):
\begin{equation} \label{gauge}
\left[ h_{a b}-\frac{1}{2}\eta _{a b}h\right]_{,b }=0.
\end{equation}
A similar gauge condition, but with one scalar component, in
Maxwell-Lorentz electrodynamics is named Lorenz
 (not from H. A. Lorentz).
If we substitute the Ricci tensor (\ref{Ricci}) and the curvature
scalar (\ref{scal}) into Einstein's equations, one obtains
\begin{equation} \label{h}
-\frac{1}{2} \Box h_{a b}+\frac{1}{4}\eta _{a b }\Box h= -
\frac{8\pi G}{c^{4}}T_{a b}
\end{equation}
We now define the gravitational potentials as
\begin{equation} \label{bar}
\overline{h}_{a b}=h_{a b}-\frac{1}{2}\eta _{a b}h
\end{equation}
and substitute equation (\ref{bar}) into equation (\ref{h}) and
rearrange, one gets
\begin{equation} \label{lin}
\Box \overline{h}_{a b}= \frac{16\pi G}{c^{4}}T_{a b}
\end{equation}
If we write out the D'Alembertian operator, we have
\begin{equation} \label{dev}
\frac{1}{c^{2}}\frac{\partial ^{2}}{\partial t^{2}}\overline{h}_{a
b}-\triangle \overline{h}_{a b}= \frac{16\pi G}{c^{4}}%
T_{a b}
\end{equation}
This is the basic equation in terms of the gravitational
potential, upon which all the analogies between electromagnetism
and gravity are based, and leads to some predictions, as for
instance, the existence of linear gravitational waves.

On the other hand, the 3+1 spacetime splitting, see [5], regards
three dimensional space as curved rather than Euclidean and its
metric $g_{ik}$ is the spatial part of the spacetime metric $g_{a
b}$. In this curved three space reside two gravitational
potentials: a ''gravitoelectric''
 Newtonian scalar potential $\Phi $, which resides in the time-time part $g_{00}$
of the space-time metric; and a ''gravitomagnetic'' vector
potential ${\bf{a}}$, which is essentially the time-space part
$g_{0j}$ of the space-time metric. The decomposition of $g_{a b}$
into $g_{ik}$, $\Phi $ and ${\bf{a}}$ is analogous to the
decomposition of the four vector electromagnetic potential $A_{a
}$ into an electric scalar potential $\Psi =A_{0}$ and a magnetic
vector potential $ {\bf{A}}=A_{j}$.

\section{Gravitomagnetic equations in terms of fields}
The analogy with Maxwell-Lorentz electrodynamics can be further
continued (see [6,3]), by writing the linear gravitational
equations in terms of first derivatives of the gravitational
potential, i.e., acceleration fields. For doing this, we first
introduce the object
\begin{equation} \label{6}
G^{abc}={\frac{1}{4}}\left(\overline{h}^{ab,c}-\overline{h}^{ac,b}
 +  \eta^{ab}\,\overline{h}^{cd}_{\;\;\;\; ,d} - \eta^{ac}\,
 \overline{h}^{bd}_{\;\;\;\; ,d}\right)
\end{equation}
 Impose now the harmonic de Donder
 gauge condition:
\begin{equation} \label{7}
\overline{h}^{ab}_{\;\;\;\; ,b}=0.
\end{equation}
From (\ref{6}) and (\ref{7}) one obtains:
\begin{equation}\label{g}
 G^{abc}={\frac{1}{4}}\left(\overline{h}^{ab,c}-\overline{h}^{ac,b}\right)
\end{equation}
Its properties are:
\begin{eqnarray} \label{10}
G^{a[bc]}   &=&G^{abc},\\ \label{8} G^{[abc]}   &=&0,          \\
\label{9} G^{d[ab,c]}&=&0,
\end{eqnarray}
where $[\cdot\cdot]$ is the antisymmetry symbol.
 Taking (\ref{1}),(\ref{6})
and (\ref{7}) into (\ref{ein}) and keeping
 only linear terms, one obtains the weak field equations in terms
 of the object $ G^{abc}$, in which only
 first derivatives of the gravitational potential appear:
\begin{equation}\label{11}
\frac{\partial G^{abc}}{\partial x^c}= -\frac{4\pi G}{c^{4}}
T^{ab}.
\end{equation}
 Introducing the
gravitoelectric newtonian scalar potential $\Phi$ and
 the gravitomagnetic vector potential ${\bf{a}}$ as
 \begin{equation}\label{g1}
\Phi = -\frac{c^{2}\overline{h}^{00}}{4}
\end{equation}
\begin{equation}\label{g2}
a^{i}=\frac{c^{2}\overline{h}^{0i}}{4}
 \qquad {{\bf{a}}}=\left(a^{1},a^{2},a^{3}\right) .
\end{equation}
Introduce new symbols, and substitute equations (\ref{g1}) and
(\ref{g2}) into equation (\ref{g}), to get the gravitoelectric
field ${\bf{g}}$ as
\begin{equation}\label{pa15}
G^{00i}=\frac{1}{4}\left(
\overline{h}^{00,i}-\overline{h}^{0i,0}\right),
\end{equation}
\begin{equation}\label{pa16}
g^{i}= c^{2}G^{00i}=-\frac{\partial \Phi }{\partial
x^{i}}-\frac{1}{c}\frac{\partial {a^{i}}}{\partial t},
\end{equation}
\begin{equation}\label{Maxg}
{{\bf{g}}}=-\nabla \Phi - \frac{1}{c}\frac{\partial \bf {a} }{%
\partial t},
\end{equation}
and the gravitomagnetic field $\bf{b}$ as
\begin{equation}\label{12}
\begin{array}{llll}
{\bf{b}}=(b^1,b^2,b^3), \hspace{5mm}& b^1=c^2G^{023}, &
b^2=c^2G^{031}, &  b^3=c^2G^{012}
\\[5mm]{\bf{b}}=\nabla\wedge {\bf{a}} &&&
c^{2}G^{0ij}=a^{i,j}-a^{j,i} .
\end{array}
\end{equation}
On the other hand, performing now the first order slow motion
approximation for the energy-momentum tensor, we assume that the
source masses involved have
 appreciable velocity $\bf{v}$ or
rotation, but neglect quadratic terms in velocity, i.e., neglect
 the stress part of the energy-momentum tensor.
Thus, the energy-momentum tensor will only have the components
\begin{equation}\label{pa5}
T^{00}=\rho c^{2}
\end{equation}
and
\begin{equation}\label{comp}
T^{0i}=\rho cv^{i}.
\end{equation}
 Put (\ref{pa16}), (\ref{12})
(\ref{pa5}) and (\ref{comp}) into (\ref{11}) and (\ref{10}), one
obtains, when the first order effects of the motion of the sources
are taken into account, the following {\it gravitomagnetic}
(Mawwell-like) equations:
\begin{eqnarray}
\nabla{{\bf{g}}}             & =&  -4\pi G \rho,             \\
\label{13} \nabla\wedge{\bf{b}} &=& -\frac{4\pi G }{c}\,\rho\,
{\bf{v}}+ \displaystyle\frac{1}{c} \frac{\partial
{{\bf{g}}}}{\partial t}, \\ \label{14} \nabla\wedge{{\bf{g}}} &=
&-\displaystyle\frac{\partial{\bf{b}}}{\partial t},\\ \label{15}
\nabla{\bf{b}} &=&0. \label{16}
\end{eqnarray}
From the four harmonic gauge conditions one obtains that, at order
$c^{-2}$, the potentials verify only the Lorenz-like one:
\begin{equation}\label{lor}
 \nabla {\bf{a}} +\frac{1}{c} \frac{\partial{\Phi}}{\partial t}=
 0.
\end{equation}
At this point it must be emphasized that in [1], the introduction
of an additional vector gauge condition was considered, which
reads
\begin{equation}
   \frac{\partial {\bf{a}}} {\partial t} = 0,
\end{equation}
as also coming from the harmonic gauge conditions. This is in our
opinion incorrect, and moreover, causes the disappearance of the
induction term in~eq.(\ref{14}).

 However, in the particular case
in which the weak gravity field is stationary, i.e., if
 the gravitational
potentials are independent of time
\begin{equation}\label{17}
  \overline{h}^{00}_{\;\;\;\; ,0}=0,  \quad \,\,\,\,  \overline{h}^{0i}_{\;\;\;\;
  ,0}=0,
\end{equation}
or equivalently, due to (\ref{g1}) and (\ref{g2}),
\begin{equation}
 \frac{\partial \Phi}{\partial t} = 0,  \quad \,\,\, \frac{\partial
 {\bf{a}}}
 {\partial t} = 0,
\end{equation}
then the gravitomagnetic equations are transformed into the
following ones
\begin{eqnarray}
\label{newgau}
 \nabla{{\bf{g}}}             & =&  -4\pi G \rho,             \\
\label{22} \nabla\wedge{\bf{b}} &=& -\frac{4\pi G }{c}\,\rho \,
\bf{v}, \\ \label{23} \nabla\wedge{{\bf{g}}} &=&0,\\ \label{24}
\nabla{\bf{b}}            &=&0.
\end{eqnarray}
and in this case the gauge condition is the Coulomb-like one
\begin{equation}\label{cou}
   \nabla {\bf{a}}= 0,
\end{equation}
  where now
${\bf{g}}$ is the Newtonian acceleration field
\begin{equation}\label{new}
{{\bf{g}}}=-\nabla \Phi.
\end{equation}
Finally, in the static case $(\overline{h}^{0i} = 0)$, the
gravitomagnetic field $\bf{b}$ is zero and one obtains the
Newtonian field equations:
\begin{equation} \nabla{{\bf{g}}}
   =-4\pi G \rho,
\end{equation}
\begin{equation}
\nabla\wedge{{\bf{g}}}=0.
\end{equation}
 Also, for a weak stationary gravity field, from the geodesic equation
 (equations of motion)
\begin{equation}\label{18}
\frac{d^2x^a}{d\tau^2}+\Gamma^a_{bc}\,\frac{dx^b}{d\tau}\frac{dx^c}{d\tau}=0,
\end{equation}
 one obtains the Lorentz-like force law, which reads
\begin{equation}\label{19}
\frac{d\bf{u}}{dt} ={\bf{g}} + \frac{4}{c} \,\bf{u}\wedge{\bf{b}}
,
\end{equation}
where $\bf{u}$ is the velocity of the test particle. Note the
factor 4 in the gravitomagnetic force term, which signals the
difference with respect to electrodynamics. Moreover, for a weak
stationary field one obtains the gravitomagnetic potential
\begin{equation}\label{20}
{\bf{a}} =-{\frac{1}{2}}\frac{\bf{S} \wedge \bf{r}}{cr^3},
\end{equation}
and the gravitomagnetic field
\begin{equation}\label{21}
{\bf{b}}=\nabla\wedge
{\bf{a}}=-{\frac{1}{2}}\frac{3\bf{n}(\bf{S}\cdot \bf{n})-\bf{S}}{c
 r^3},
\end{equation}
where $\bf{S}$ is the intrinsic angular momentum of the source and
$\bf{n}$ is the unit position vector. These equations are
analogous to the electromagnetic ones, changing the magnetic
dipole moment by minus half the angular momentum. They are used in
the GP-B gyroscope and LAGEOS III experiments (see [7]), to obtain
the precession of test bodies due to stationary gravitomagnetic
$\bf{b}$ field generated by the rotation of the Earth mass.

\section{Conclusion}
The gravitational potentials $\Phi$ and ${\bf{a}}$ can be time
dependent, for instance, due to changes in time of the mass,
intrinsic angular momentum or distance of(to) the source. In this
work, we have showed that the gravitomagnetic equations, with an
induction term in the Faraday-like one, must be applied in this
case, when the linear and first order slow motions approximations
of the gravitational field are considered.

\section{ACKNOWLEDGEMENTS}
 I am grateful to B. Mashhoon for correspondence
about the subject and to A. San Miguel and F. Vicente for
discussions on this and (un)related topics. This work has been
partially
 supported by the spanish research
projects VA61/98, VA34/99 of Junta de Castilla y Le\'on and
C.I.C.Y.T. PB97-0487.

\section{REFERENCES}

[1] Harris, E. G., {\it Am. J. Phys.} {\bf59} (1991) 421.\\
\noindent [2] Ohanian H. C., Ruffini R., {\em Gravitation and
Spacetime},
 (W.W.Norton \& Company) 1994, pp.~163.\\
 \noindent
[3] Pascual-S\'anchez~J.-F. in {\em The Lense-Thirring effect},
 to be edited by Fang L.Z. and Ruffini R., (World Scientific).\\
 \noindent
[4] Wald R.M., {\em General Relativity}, (The University of
Chicago Press) 1984.\\ \noindent
 [5] Jantzen~R.~T., Carini~P. and Bini~D.,
{\it Ann. of Phys.} {\bf215}
(1992) 1.\\
 \noindent
 [6] Peng~H., {\it Gen. Rel.
Grav.} {\bf15} (1983)725.\\
 \noindent
 [7] Ciufolini I. and Wheeler
J. A., {\em Gravitation and Inertia}, Princeton Series in Physics,
pp.~315-383, 1993.

\end{document}